\documentclass[12pt]{article}
\usepackage{psfig}
\usepackage{graphicx}

\begin{document}

{Kinematics and Physics of Celestial Bodies Vol. 21, No. 1, pp. 1-16, 2005} \\ \\ \\ \\

\begin{center}

\bigskip

{\LARGE \bf Triplets of Galaxies in the Local Supercluster. I. Kinematic and Virial Parameters} \\

{\small I. B. Vavilova$^{1}$, V. E.Karachentseva$^{1}$, D. I. Makarov$^{2}$ and O. V. Melnyk$^{1}$}  \\

{\em $^{1}$Astronomical Observatory, Kiev National University, Observatorna str., 3,
Kiev, 304053 Ukraine \\
melnykov@observ.univ.kiev.ua \\
$^{2}$Special Astrophysical Observatory, Russian Academy of Sciences,
Nizhnii Arkhyz, Karachai-Cherkessian Republic, 357147 Russia}\\
\end{center}

{\large \bf Abstract}\\
We considered kinematic and virial parameters of a sample of galaxy triplets from the Local Supercluster (LS). The galaxies were selected from the LEDA database, and their radial velocities are no more than 3100 km/s. The sample contains 176 systems selected according to the Karachentsev criterion. We calculated the following median parameters of the LS triplets: rms velocity of galaxies with respect to triplet center, $S_{v} =$ 30 km/s; harmonic mean radius, $R_{h} = 160$ kpc; virial mass, $M_{vir}= 3.6\cdot 10^{11}M_{\odot}$; and mass-to-light ratio, $M_{vir}/L = 35M_{\odot}/L_{\odot}$. Although various investigators used different approaches to compile samples of triple sys­tems, general properties of the triplets in all the samples we considered agree rather satisfac­torily. The LS triplets have the least mass-to-light ratio as compared to other galaxy triplet samples.

{\bf Key words : galaxies, groups of galaxies, Local Supercluster.}

\bigskip

\section{Introduction}
Triplets of galaxies represent poorly populated galaxy groups. Of special interest for modern astrophysics is to study kinematic and dynamical parameters of triple systems with the aim to clarify the role of triplets in the evolution of galaxy groups, to estimate the amount of dark matter on small scales, and to make a comparison to the same parameters of single galaxies, double galaxies, and rich groups and clusters.
The mass-to-light ratio $M/L$ monotonically increases with the population of galaxy groups, as was first demonstrated by Karachentsev in 1966. Since the end of the 1970s, the amount of dark matter in galaxy groups, in particular, in triplets (Karachentsev et al. 1989, Trofimov, Chernin 1995) and in larger groups (Tully 1987, 2005) was estimated. Some evidence for the existence of dark matter in the halos of single galaxies as well as in the space between the galaxies from the same group or cluster was found in (Karachentsev et al. 1989, Trofimov, Chernin 1995, Tully 2005). So, the virial excess observed in galaxy groups can be explained by an increase in the number of single galaxy halos with the population of the group as well as by an increase of the size of the group with the distance between the galaxies (Karachentsev et al. 1989).

The virial theorem is valid for closed systems in the state of dynamic equilibrium only, provided that parameters of the system are averaged over the system lifetime. It is difficult to determine whether this condition actually holds, therefore, we can only make an assumption of its validity when calculating the virial mass for galaxy groups and clusters. The three-body problem is much more complicated than the two-body problem. Several solutions demand additional approximations (see, e.g., study (Orlov et al. 2002) for the collinear solution for three equal masses and the equilateral triangle solution for three different masses). Recent numerical simulation results improved our knowledge about the behavior of ensembles of three-body systems rather than about the individual behavior of these systems. Systems of triple galaxies are characterized by unclosed orbits, and this throws doubt on the correctness of using the virial theorem to determine the mass-to-light ratios for particular triplets (Karachentseva, Karachentsev 1982). Since the dynamical nonstationarity of triplets as well as projection effects introduce large errors in the virial mass estimates obtained from numerical simulation, the virial masses of individual triplets cannot be determined (Chernin, Mikkola 1991), and the median mass-to-light ratio for a statistical ensemble of triplets was found to be $M_{vir}/L = 70M_{\odot}/L_{\odot}$. Studies (Karachentsev et al. 1989, Karachentseva, Karachentsev 2000) independently gave the estimate $\{M_{vir}/L\}_{1/2}  = 20-60 M_{\odot}/L_{\odot}$, which is at least four times as large as the mass-to-light ratio for single and double galaxies.	
The mass-to-light ratio for groups of more than five galaxies was estimated at $\{M_{vir}/L\}_{1/2}  = 94 M_{\odot}/L_{\odot}$ (Tully, 2005).	

The aim of this study is to determine kinematic and virial parameters of the triple galaxy systems	
located in the Local Supercluster (LS) and to make a comparison to the same parameters for other triplet	
samples. The selection criteria and the list of the LS triplets are described below.	

There are two special lists of isolated triplets compiled by I. D. Karachentsev and V. E. Karachentseva	
on the basis of the Palomar Sky Survey POSS-I and ESO/SERC survey, namely, the list of Northern triplets	
(Karachentseva et al. 1979) (it is limited by the apparent magnitude $m \le 15.7$ (Karachentsev et al. 1989, Karachentseva, Karachentsev 1982, Karachentseva et al. 1979, 1987,  Trofimov, Chernin 1995) and the list of Southern triplets (Karachentseva, Karachentsev 2000) (it is limited by the angular diameter $a \ge 1'$). The zones with $\mid b\mid \le 20^{\circ}$ were excluded from the both samples.

The procedure used to select the Northern and Southern triplets was based oniy on the data about the	
angular dimensions of galaxies and relative distances between them. The galaxies which were supposed to	
form a triplet were considered together with their "significant" neighbors located at least three times as far	
from each other as the galaxies in the triplet. Taking "significant" neighbors into account could, to some	
extent, cancel the influence of the near and far background. The selection principle was to reveal an isolated	
triple of galaxies, which is clearly distinguished in the projection on the sky (Karachentseva et al. 1979, Karachentseva, Karachentsev 2000). Note that radial velocities were not analyzed, and consequently a system qualified as a triplet may turn out to be isolated in the projection only.	

In addition to samples (Karachentseva et al. 1979, Karachentseva, Karachentsev 2000), we used two other galaxy triplet samples for comparison: the list of Wide triplets (Trofimov, Chernin 1995) and the list composed mainly of the triplets from Tully's catalog (Tully 1987). Wide triplets were selected	
from two catalogs of galaxy groups: catalog (Geller, Huchra 1983) for the northern sky (the sample is limited by the apparent	
magnitude $m \le 14.5$ and by the zones $b_{II} \ge 40^{\circ}$, $\delta\ \ge 0^{\circ}$ and $b_{II} \le 40^{\circ}$, $\delta\ \ge -2.5^{\circ}$ ) and catalog (Maia et al.) for the southern sky ($b_{II} \ge -30^{\circ}$, $\delta\ \le -17.5^{\circ}$ ). The sample was compiled with the constraints that the radial velocities of galaxies should not exceed 12 000 km/s and the radial velocities of the galaxies from the same triplet should differ by no more than 600 km/s. Note, however, that the sample of Wide triplets did not include all the triplets from these catalogs. The systems which looked like triplets in the projection only were excluded using the criterion (Anosova 1987) for a triple system to be isolated and bound. Tully's catalog (http://www.anzwers.org/free/ universe/galaclus.html) embraces the whole sky (except the zone$\mid d\mid \le 20^{\circ}$) and covers radial velocities up to 3000 km/s. An hierarchical approach was used to select the groups of galaxies (Tully 1987, 2005). The galaxies were considered in pairs to select such a pair for which the quantity $LR_{ij}^{-2}$ is maximum ($L$ being luminosity of the brightest component and $R_{ij}$ distance between the components). This pair was then returned to the catalog as a single object with combined luminosity. The selection procedure was repeated until the physically bound pairs of galaxies were all sorted out. (A group of galaxies was assumed to be physically bound if $t_{x}H_{0}<< 1$, where $t_{x} \sim R_{I}/V_{p}$ is the crossing time, $R_{I}$ the inner radius of the system, and $V_{p}$ the velocity dispersion.)

\section{Sample of LS Galaxy Triplets}

The sample of the LS galaxies with $V_{LG} < 3100$ km/s, $V_{LG}$ being the radial velocity corrected for the Sun's motion according to formula (2), was composed on the basis of the LEDA2003 database (Lyon-Meudon Extragalactic Database). This LEDA version was compiled from various galaxy catalogs and surveys, and it contains more than three million objects. Although Paturel and his group made a lot of effort to refine the database, it still contains unconfirmed (and sometimes erroneous) compilation data, e.g., coincident records and missing parameters (type, magnitude, angular dimensions, etc.). In view of this fact, we examined the objects with close coordinates to eliminate coincident records. To exclude extragalactic objects and false identifications, the objects with radial velocities from -900 km/s to 1000 km/s were considered each sepa­rately. We also excluded the galaxies with large measurement errors (greater than 150 km/s) in radial velocities. By comparing the radial velocities measured in the radio frequency range to the same data obtained in the optical range, we eliminated doubtful records and typographic errors. To check the sample, we used the NED database (NASA Extragalactic Database), the catalogs of dwarf galaxies with low surface brightness, the POSS-I and ESO/SERC charts. As a result, we obtained a reasonably refined sample which contains about 7000 galaxies with $V_{LG} < 3100$ km/s and $\mid b\mid > 10^{\circ}$).

The next step was to single out physically bound groups of galaxies. To this end, we used the method (Makarov, Karachentsev 2000, Tikhonov, Makarov 2003) based on the Karachentsev algorithm (Karachentsev 1994). This method was used earlier by Makarov and Karachentsev (2000) to compile the catalog of LS groups (hereafter referred to as MK2000 catalog) on the basis of LEDA1996. Any criterion for selecting a physically bound group of galaxies (in particular, a triplet) implies that ($i$) the group should be isolated (it should not be a part of any larger group or cluster) and ($ii$) the group may not contain "optically false" components belonging to the near or far background. Karachentsev's method (1994) allowed us to exclude the majority of "optically false" galaxies. The review of other selection methods can be found in (Karachentsev 1994, Vavilova 1998).

\begin{figure}[ht]
\centerline{\includegraphics[angle=0, width=14cm]{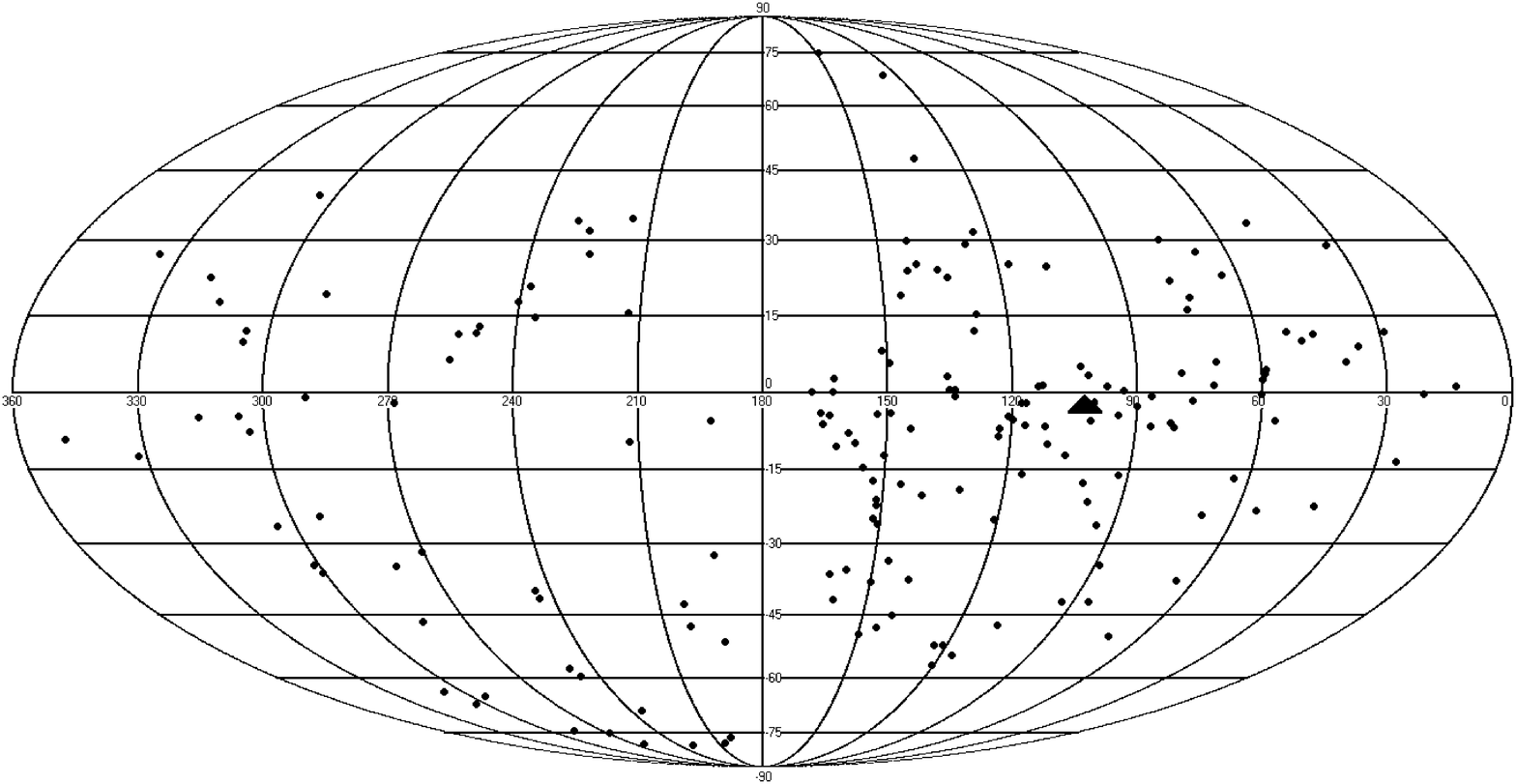}}
\caption{Distribution of LS triplet centers over supergalactic coordinates. Triangle - Virgo cluster center.}
\end{figure}

Karachentsev's method (1994) allows the observable parameters of galaxies to be analyzed most thoroughly. At the first stage, all the possible combinations of galaxy pairs are considered. A pair is assumed to be physical when its total energy is negative. Very wide false pairs with small radial velocity differences are excluded by applying the constraint that the projected distance between the galaxies should be less than the radius of the collapsing sphere (the so-called zero-velocity surface) (Sandage 1986). At the second stage, the pairs are all combined in groups: the pairs with the same main galaxy form a single group. When the main galaxy of some group is also a member of another group, these two groups are merged, and the most massive galaxy is further treated as a new main galaxy. In the groups selected, we additionally analyzed the objects with very small relative projected distances. As a result, some pairs and groups turned out to be single galaxies. In most cases, these were sufficiently close galaxies in which the radial velocities of bright H II regions were measured. For such galaxies, different parts of the same galaxy might erroneously be included in LEDA under different numbers.

So, we selected 3778 LS galaxies that form groups of 2 to 82 galaxies. In particular, 14 percent of these galaxies constitute 176 triplets. Figure 1 demonstrates the positions of the centers of these LS triplets. For comparison, the MK2000 catalog comprises 156 triplets whose overall population constitutes 13 percent of the total number of galaxies ($N = 3472$) in groups. Therefore, the percentage of triplets in our new sample of LS groups is almost the same as in the MK2000 sample. Our sample includes 71 triplets (40 percent of 176 triplets in total), which are also found with all three their components in the MK2000 sample. There are also 48 triplets (26 percent) comprising new galaxies that are absent in the MK2000 catalog. Other triplets were formed from the galaxies redistributed in groups as compared to the MK2000 sample (for the most part, these are the triplets separated from more populated galaxy groups). Such a change in the population of triplets took place because ($i$) the total number of galaxies in the new version of LEDA was increased, ($ii$) doubtful velocity measurements and "false" galaxies were excluded, and ($iii$) clusterization parameters were slightly changed. Using new data for apparent magnitude, radial velocity, and other galaxy parameters led to the inclusion of some galaxies in groups as well as to the exclusion of some other galaxies from these groups. By comparing the MK2000 sample and our new sample compiled from LEDA2003, we can examine the sensitivity of the criterion used to select bound groups of galaxies and analyze the influence of this criterion on the parameters of galaxy groups.

\section{Physical Parameters of LS Galaxy Triplets}

To calculate kinematic and virial parameters of the LS triplets, we used the following formulae (each triplet was considered as an indivisible dynamical system (Karachentsev et al. 1989, Karachentseva, Karachentsev 2000): 

- mean corrected radial velocity of the triplet as a whole:
\begin{equation}
\label{trivial}
\langle V_{LG}\rangle = \sum_{k=1}^3 V_{LG}^k/3.                                                            
\end{equation}

The radial velocities of triplet components were taken from LEDA and NED. The heliocentric radial velocities of galaxies were corrected for the Sun's motion:
\begin{equation}
\label{trivial}
V_{LG}=V_{h}+316[\cos b \cos (-4^{\circ})\cos (l - 93^{\circ})+\sin(b)\sin(-4^{\circ})],        
\end{equation}
where $l$ and $b$ are the galactic coordinates of galaxies. The solar apex parameters were taken from (Karachentsev, Makarov 1996)

- rms velocity of galaxies with respect to the triplet center:
\begin{equation}
\label{trivial}
S_{v}=\left[\frac{1}{3} \sum_{k=1}^3(V_{LG}^k-\langle V_{LG}\rangle)^{2} \right]^{1/2}
\end{equation}
- relative linear distances between the triplet components in the projection:
\begin{equation}
\label{trivial}
R_{ik}=X_{ik}\langle V_{LG}\rangle H_{0}^{-1},
\end{equation}
where $X_{ik}$ are the relative angular distances; 

-mean and harmonic mean radii of the triplet:
\begin{equation}
\label{trivial}
\langle R_{ik}\rangle = \frac{1}{3}\sum_{i,k} R_{ik},
\end{equation}

\begin{equation}
\label{trivial}
R_{h}=\left[\frac{1}{3} \sum_{i,k} R_{ik}^{-1} \right]^{-1}
\end{equation}
- dimensionless crossing time expressed in units of the cosmological time $H_{0}^{-1}$ and calculated with mean radus (5) and harmonic mean radius (6) of the triplet:

\begin{equation}
\label{trivial}
\tau _{\langle R\rangle}=2H_{0} \langle R\rangle /S_{v}
\end{equation}

\begin{equation}
\label{trivial}
\tau =2H_{0} R_{h} /S_{v}
\end{equation}

- virial mass of the triplet as a whole (Karachentseva, Karachentsev 1982) calculated with the mean and harmonic mean radii:

\begin{equation}
\label{trivial}
M_{vir, \langle R\rangle} = 3\pi N(N-1)^{-1}G^{-1}S_{v}^{2} \langle R\rangle,  
\end{equation}

\begin{equation}
\label{trivial}
M_{vir} = 3\pi N(N-1)^{-1}G^{-1}S_{v}^{2}R_{h},  N=3
\end{equation}
where $G$ is the gravitational constant;

- luminosity of each galaxy in solar units (Karachentsev et al. 1999):

\begin{equation}
\label{trivial}
\log L = 2\log (V_{LG}/H_{0})-0.4[B_{T}-(a_{g}+a_{i})]+12.16
\end{equation}

where $B_{T}$  is the total $B$ magnitude, $a_{g}$ is the correction for the absorption in our Galaxy (Schlegel et al. 1998), and $a_{i}$ is the
correction for the internal galactic absorption. The parameters $B_{T}$, $a_{g}$, and $a_{i}$, were taken from LEDA;

- unbiased estimate of the mass-to-light ratio with the measurement errors of radial velocities taken into
account:

\begin{equation}
\label{trivial}
f^{c}=f[1-2\sigma _{v}^2/3S_{v}^2]
\end{equation}

where $\sigma _{v}$ is the rms error of the radial velocities of triplet components and $f=M_{vir}\left[ L_{i} \right] ^{-1}$, $i=$ 1, 2, 3.  We also
calculated the unbiased estimate of mass-to-light ratio $f_{\langle R\rangle}^{c}$, with the virial mass determined by formula (9). 

The Hubble constant $H_{0}$ was assumed to be equal to 75 km/ (s $\cdot$ Mpc). 

Table 1 gives the following triplet parameters:

1. Number in order of increasing right ascension.

2. Names of galaxies according to the main LEDA specification.

3. Mean corrected radial velocity calculated in km/s with formulae (1) and (2).

4. Rms velocity of galaxies (in km/s) with respect to triplet center, formula (3).

5. Mean radius (5) in kpc.

6. Harmonic mean radius (6) in kpc.

7. Dimensionless crossing time (7) calculated with mean radius.

8. Dimensionless crossing time (8) calculated with harmonic mean radius.

9. Logarithm of virial mass expressed in $M_{\odot}$, formula (9).

10. Logarithm of virial mass expressed in $M_{\odot}$, formula (10).

11. Logarithm of luminosity expressed in $L_{\odot}$, formula (11).

12. Mass-to-light ratio $f_{\langle R\rangle}$.

13. Unbiased estimate of the mass-to-light ratio $f_{\langle R\rangle}^{c}$, with the virial mass calculated by formula (9).

14. Mass-to-light ratio $f$.

15. Unbiased estimate of the mass-to-light ratio $f^{c}$, with the virial mass calculated by formula (10).

The median mass-to-light ratios calculated with the mean and harmonic mean radii are $f_{\langle R\rangle}=49 f_{\odot}$
and $f = 35  f_{\odot}$, respectively. The median unbiased estimates are $f_{\langle R\rangle}^{c}=37 f_{\odot}$
 and $f^{c} = 27  f_{\odot}$, with all the triplets taken into account. When 18 triplets for which the measurement error of radial velocity exceeds the rms velocity are excluded, we have $f_{\langle R\rangle}^{c}=42 f_{\odot}$
 and $f^{c} = 33  f_{\odot}$. Thus, the measurement errors of radial velocities only slightly affect the mass-to-light ratios.

Let us examine some relationships between various parameters listed in Table 1.

Figure 2 demonstrates the distribution of the LS triplets over the mean and harmonic mean radii. The median radii are $\langle R\rangle= 259$ kpc and $R_{h}=$ 160 kpc. The mean radii of triplets 12, 55, 57, 84, 101, 116, 120, 136, and 139 as well as the harmonic mean radii of triplets 101 and 120 are greater than 600 kpc.

\begin{figure}[p]
\centerline{\includegraphics[angle=0, width=10cm]{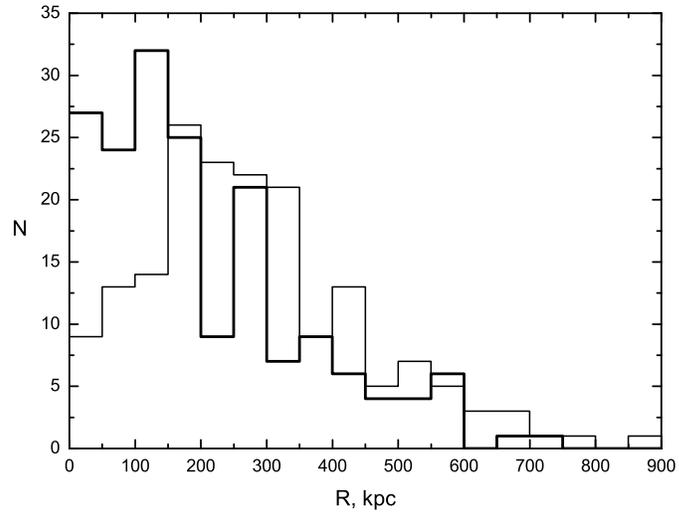}}
\caption{Distribution of LS triplets over the mean $\langle R\rangle$ (thin line) and harmonic mean $R_{h}$ (thick line) radii.}
\end{figure}

Figure 3 presents the distribution of the LS triplets over the rms velocity, the median rms velocity being $S_{v}=30$ km/s. Triplet 13 has the largest velocity dispersion (246 km/s).

\begin{figure}[p]
\centerline{\includegraphics[angle=0, width=10cm]{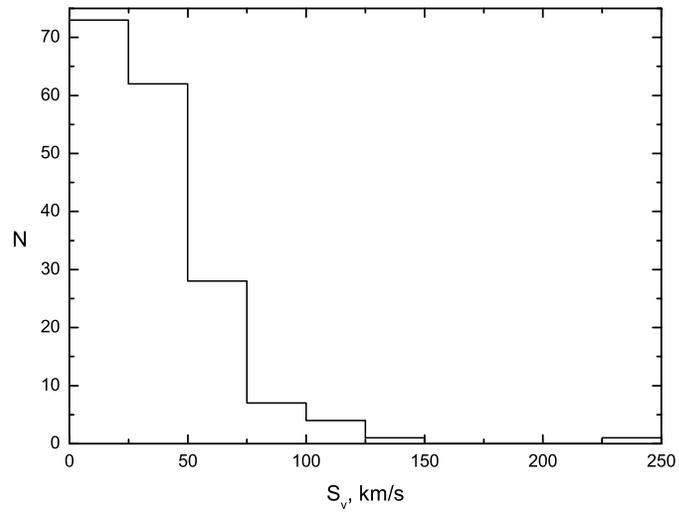}}
\caption{Distribution of LS triplets over the rms velocity of triplet galaxies.}
\end{figure}

Figure 4 shows the distribution of the LS triplets over the virial mass. The median virial masses calculated with the mean and harmonic mean radii are $M_{vir, \langle R\rangle} = 5.9\cdot 10^{11} M_{\odot}$ and $M_{vir}= 3.6\cdot10^{11}M_{\odot}$. Triplets 13 and 66 have the largest virial masses $M_{vir}$, namely, $1.55\cdot10^{13}M_{\odot}$ and $1.2\cdot10^{13}M_{\odot}$. The virial masses of triplets 5, 39, 94, and 98 are lower than $10^{11}M_{\odot}$.

\begin{figure}[t]
\centerline{\includegraphics[angle=0, width=10cm]{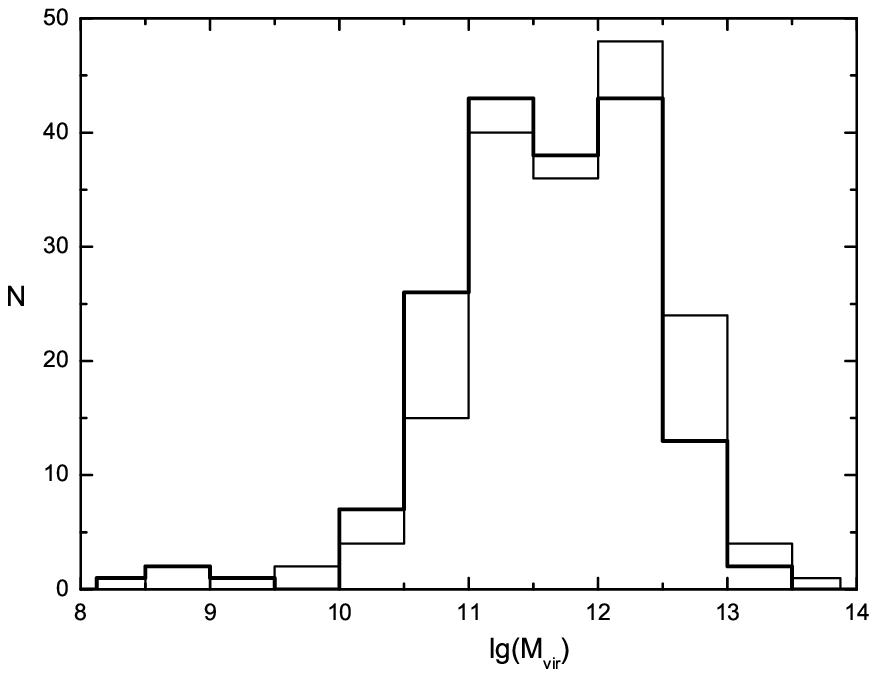}}
\caption{Distribution of LS triplets over the virial mass of a triplet as a whole in units of $M_{\odot}$: estimates based on the mean radius $M_{vir, \langle R\rangle}$ - thin line, estimates based on the harmonic mean radius $M_{vir}$  - thick line.}
\end{figure}

Triplets 86, 89, 94, 98, 99, 101, 105, and 109 lie in the Virgo cluster region ($\alpha  = 11^{h}50^{m}\div12^{h}30^{m}$ and
$\delta  = -2.5^{\circ}\div20^{\circ}$). This region is characterized by large tidal forces and internal motions of galaxies (their peculiar velocities were ignored). The parameters of some of these triplets (e.g., of triplets 101, 94, and 98) significantly differ from median estimates (see Table 1 and Figs 2 and 4).
The virial mass, crossing time, and mass-to-light ratio depend on the formula used to estimate typical triplet radii (mean radii or harmonic mean radii). The median virial parameters calculated with the mean radius are mostly larger than the same parameters calculated with the harmonic mean radius (e.g., $\tau _{\langle R\rangle}= 1.31$ and $\tau = 0.87$).

\section{Comparing the Parameters of the LS Triplets \break and Triplets From Other Samples}

Let us consider various triplet ensembles and calculate median parameters of triplets. Since LEDA and NED have recently provided new data on radial velocities, magnitudes, and other parameters of galaxies, we up­dated the Southern, Wide, and Tully triplet samples. The data for the radial velocities of the Northern triplets were taken from (Karachentseva et. al 1987), in view of the corresponding research program conducted at the 6-m telescope of the Special Astrophysical Observatory, RAS. We used formulae (1) - (11) to calculate the median triplet para­meters for all the samples (Table 2). The virial mass, crossing time, and mass-to-light ratio were calculated with the harmonic mean radius (6).

\newpage

Table 2. Median Parameters of Galaxy Triplets from Various Samples.

\begin{tabular}{|l|l|l|l|l|l|l|l|}
\hline
Triplet sample & N & $S_{v}$, km/s & $R_{h}$, kpc & $\tau $, $1/H_{0}$ & $\lg M_{vir}$ & $N^{*}$ & $M_{vir}/L$\\
\hline
LS&176 & 30 & 160 & 0.87 & 11.56 & 176& 35\\
\hline
LS, $\tau <1$ &95 & 39 & 103 & 0.34 & 11.73 & 95 & 45\\
\hline
LS, $\tau <0.2$ &29 & 57 & 46 & 0.12 & 11.79 & 29 & 39\\
\hline
MK2000 &156 & 33 & 178 & 0.91 & 11.72 & 153 & 35\\
\hline
Northern&83 & 135 & 70 & 0.04 & 12.65 & 82& 103\\
\hline
Southern$^{**}$ &48 & 170& 69 & 0.04 & 12.97 & 46 & 112\\
\hline
Wide&37 & 66 & 396 & 0.66 & 12.65 & 37 & 173\\
\hline
Tully&56 & 56 & 269 & 0.66 & 12.42 & 56 & 111\\
\hline
\end{tabular}

$^{*}$ N is the number of triplets for which we could calculate the ratio $M_{vir}/L$ (there are galaxies in LEDA with no apparent magnitude available).

$^{**}$  We considered 48 out of 76 Southern triplets for which the radial velocities of all the components are known. \\ \\

The LS triplets have the least rms velocity, the largest crossing time, the least virial mass, and the least mass-to-light ratio, on the average. As would be expected, the median parameters of the LS triplets are close to the median parameters of the MK2000 triplets. By imposing constraints on the crossing time ($\tau  < 1$ and $\tau  < 0.2$) in the LS sample, we found that the median rms velocity increases and the harmonic mean radius drops as the crossing time decreases, i.e., the system is closer to the virial equilibrium at smaller $\tau$ and the number of crossings is greater for the triplet galaxies. However, we found no effect of crossing time on the mass-to-light ratio. The Northern and Southern triplets have close parameters, and this seems to be a result of similar selection criteria. Wide triplets have the largest harmonic mean radius and, therefore, the largest mass-to-light ratio $M_{vir}/L$. Tully and Wide triplets can be contrasted to the Northern and Southern triplets: the velocity dispersions of Tully and Wide triplets are 2-3 times as small and the harmonic mean radii are 5-6 times as large. This distinction can be imagined in the following way: Tully and Wide triplets are larger in their size on the celestial sphere, whereas the Northern and Southern triplets are more extended along the line of sight. This effect is probably due to different selection criteria used to compile these samples.

Figure 5 shows the velocity dispersion as a function of harmonic mean radius for the LS triplets, Tully triplets, Wide triplets, and Northern and Southern triplets (we limited the rms velocity to 300 km/s in the last two samples). The LS and Tully triplets have the least scatter of data. This may indicate that optical members do not significantly affect these samples, and the mass-to-light ratios of these triplets are not considerably overestimated.

By imposing the constraint $S_{v} < 300$ km/s on the rms velocity, we made an attempt to exclude optical systems from the Northern, Southern, and Wide triplet samples. The median virial masses and mass-to-light ratios of the limited Northern and Southern triplets are comparable to the same parameters of the LS triplets (Table 3). \\ \\

Table 3. Median Virial Masses and Mass-to-Light Ratios (in Solar Units) for Various Triplet Samples.

\begin{tabular}{|l|l|l|l|}
\hline
Triplet sample & N & $M_{vir}$ & $M_{vir}/L$\\
\hline
LS&176 & $3.60\cdot 10^{11}$ & 35\\
\hline
Northern&53 & $2.09\cdot10^{12}$ & 53\\
\hline
Southern&30 & $1.75\cdot10^{12}$ & 32\\
\hline
Wide&35 & $3.25\cdot10^{12}$ & 158\\
\hline
Tully&56 & $2.60\cdot10^{12}$ & 111\\
\hline
\end{tabular}\\ \\

\begin{figure}[t]
\begin{tabular}{ll}
\includegraphics[angle=0, width=0.4\textwidth]{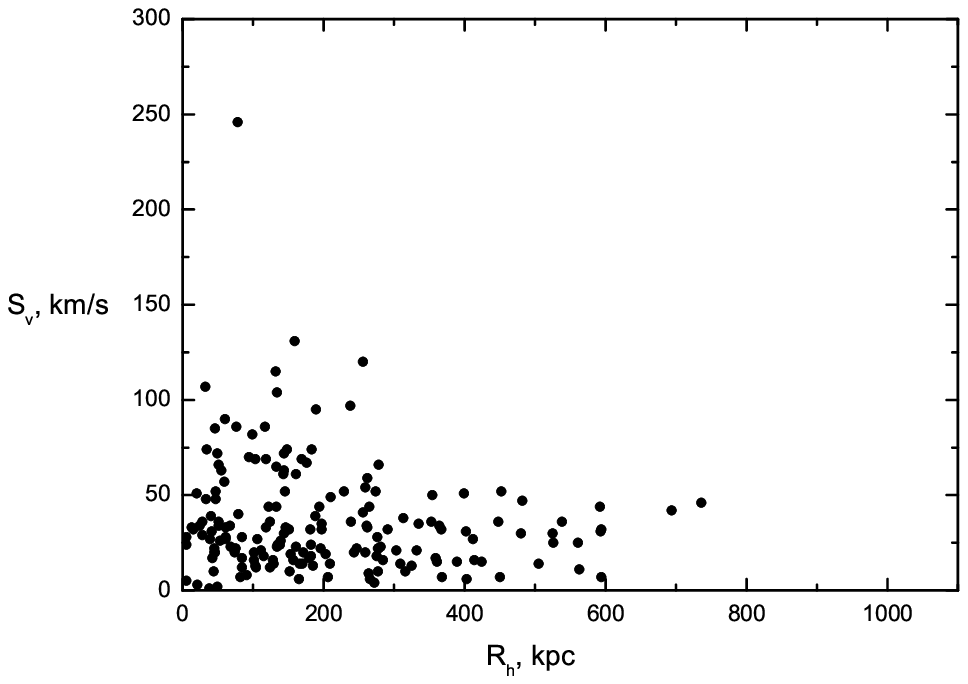} &
\includegraphics[angle=0, width=0.4\textwidth]{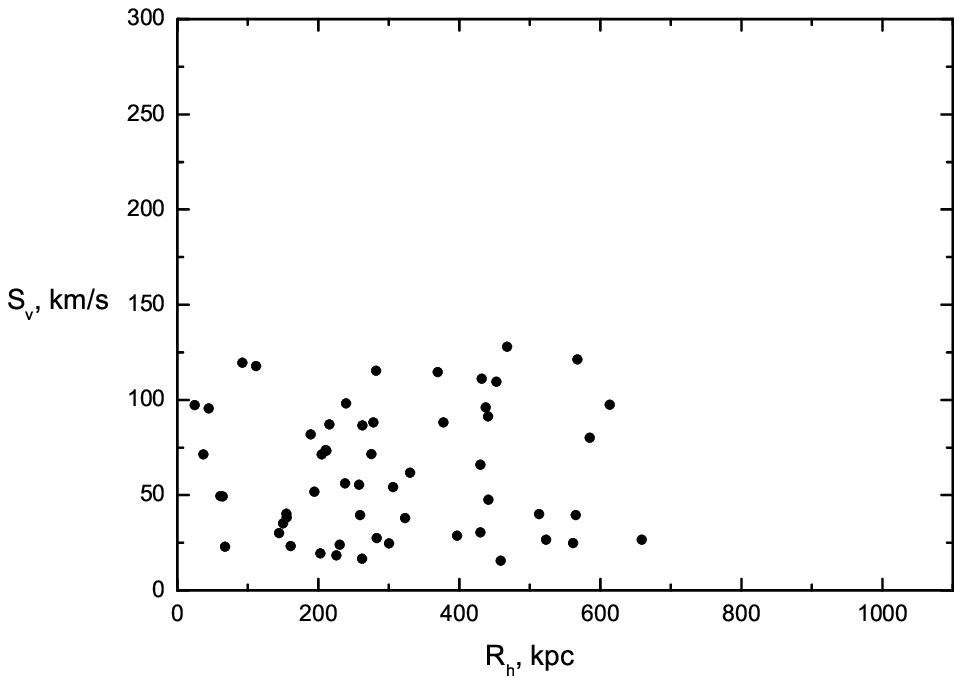} \\
\includegraphics[angle=0, width=0.4\textwidth]{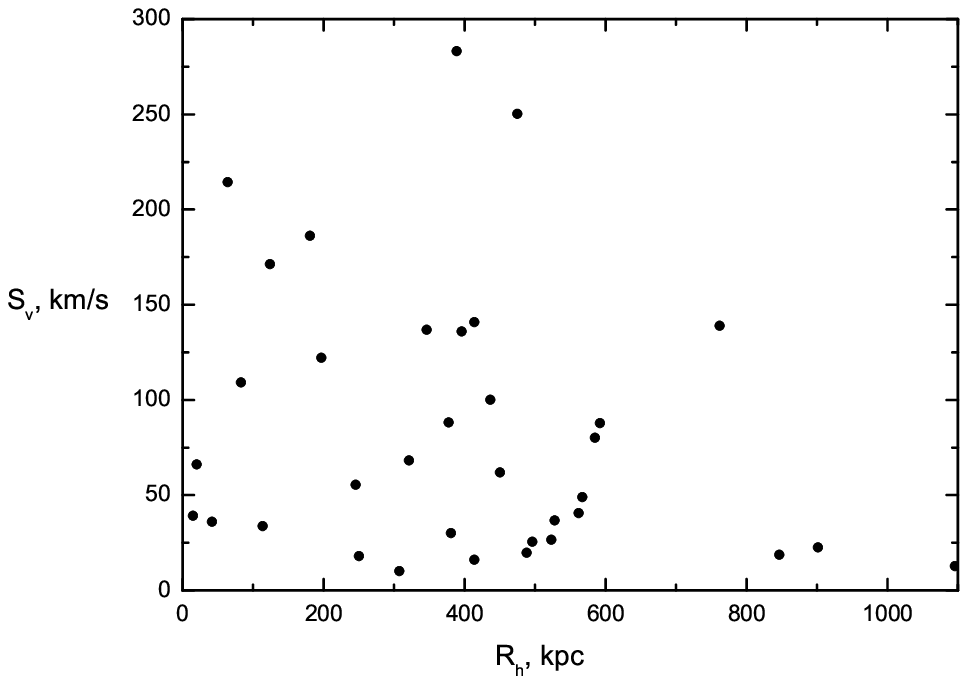} &
\includegraphics[angle=0, width=0.4\textwidth]{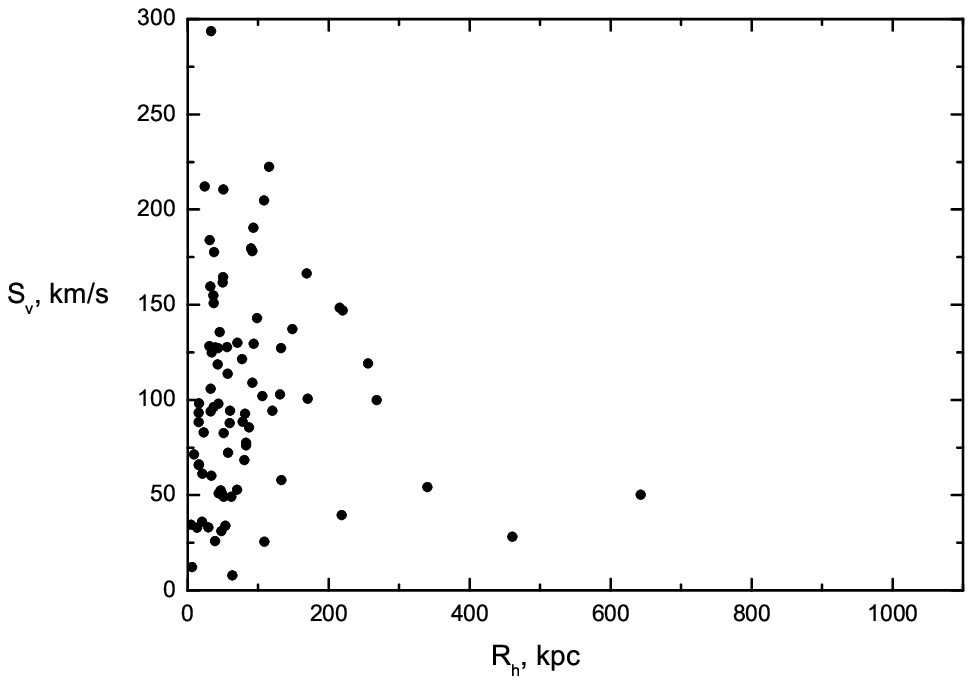}
\end{tabular}
\caption{Rms velocity versus harmonic mean radius: left top) LS triplets ($N = 176$), right top) Tully triplets ($N= 56$), left bottom) Wide triplets ($N=35$, $S_{v} < 300$ km/s), right bottom) combined sample of the Northern and Southern triplets ($N = 83$, $S_{v} < 300$ km/s). }
\end{figure}

The distinction between the Northern (Compact) and Wide triplets was discussed in \break (Karachentsev et al. 1989, Trofimov, Chernin 1995). Compact triplets have already made several revolutions about their center of mass in the course of evolution (such triplets are characterized by a small crossing time), while wide triplets are younger dynamical systems and are far from the state of virial equilibrium (such systems have large mass-to-light ratios).

The same parameters for all five triplet samples turned out to be different, and this can be explained by different selection criteria. Only two LS triplets entirely belong to the Northern and Southern triplet samples, and there are four triplets with their two components presented in both samples. There are five common triplets in the LS and Tully samples, and at least five other triplets coincide in their two components. None of the LS triplets coincides with any of the Wide triplets, and 17 Wide triplets are found (entirely or component by component) in more populated LS galaxy groups. The Wide triplets differ from the LS, Northern, and Southern triplets in their typical size (up to 1 Mpc). We can also suppose that the Wide and Tully triplets are either groups of three bright components with some dwarf satellites missed or fragments of more populated groups.

\section{Conclusion}
Based on the LEDA2003 database, we compiled a new sample of triplets of the LS galaxies with $V_{LG} < 3100$ km/s. The sample was exhaustively processed to eliminate doubtful objects. To select bound groups of galaxies, we used algorithm (Karachentsev 1994). The resulting sample contains 176 triplets which comprise 14 percent of the total number ($N=3778$) of the galaxies included in groups.

We analyzed the kinematic and virial parameters of the LS triplets and compared them to the same parameters of the Northern, Southern, Wide, and Tully triplets. In spite of different selection criteria used to compile these samples, general properties of triplets in all the samples agree quite satisfactorily. The fact that we obtained different kinematic and virial parameters for different samples can be explained by selection effects.

Comparing the biased and unbiased estimates of median mass-to-light ratios, we found that the measurement errors of radial velocities have only a slight effect on the mass-to-light ratio. The virial median parameters of LS triplets calculated with the mean radius are mostly larger than the same parameters calculated with the harmonic mean radius: ($f_{\langle R\rangle}=49 f_{\odot}$
and $f = 35  f_{\odot}$; $M_{vir, \langle R\rangle} = 5.9\cdot 10^{11} M_{\odot}$ and $M_{vir}= 3.6\cdot10^{11}M_{\odot}$; $\tau _{\langle R\rangle}= 1.31$ and $\tau = 0.87$).

\begin{figure}[t]
\centerline{\includegraphics[angle=0, width=10cm]{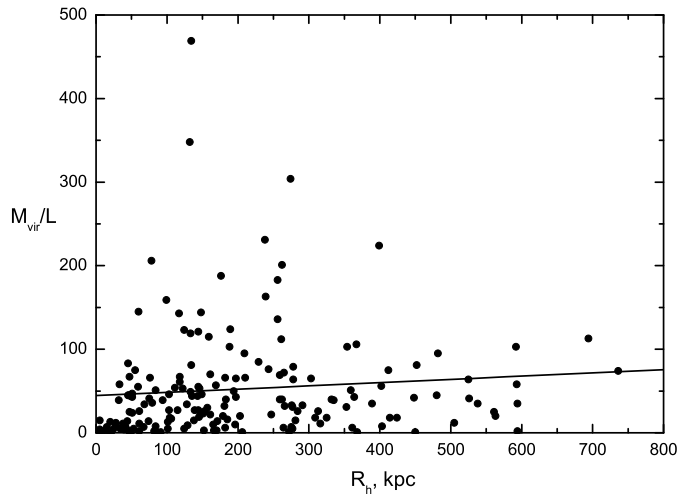}}
\caption{Mass-to-light ratio versus harmonic mean radius.}
\end{figure}

As for the problem of hidden mass, the LS triplets have the least mass-to-light ratio as compared to triplets from other samples (Tables 2 and 3). However, this mass-to-light ratio is at least 5-7 times as large as the mass-to-light ratio for individual galaxies (Karachentseva, Karachentsev 1982, Karachentsev et al. 1989). At the same time, the mass-to-light ratio for the LS triplets does not grow with their harmonic mean radius (Fig. 6). On the one hand, this may imply that the spatial scale of triplets (about 150 kpc) is too small to make definite conclusions about the existence of dark matter in the space between the galaxies. On the other hand, dark matter can predominantly be located in the halos of single galaxies, as can be deduced, e.g., from the existence of plateaus in the rotation curves of spiral galaxies, from the dynamics of stars in elliptical galaxies, and from the existence of galactic X-ray halos.

This study is the first stage in investigating galaxy groups in the Local Supercluster. Analysis of poorly populated groups of galaxies (triplets) is an important step to study more populated galaxy groups.

To investigate the LS triplets more thoroughly, we are going to study their configurational properties, to take into consideration active galaxies and X-ray sources, to estimate separately the masses of triplet components, and to analyze the effect of the morphological type of galaxies on kinematic and virial parameters of triplets.

\bigskip

{\bf ACKNOWLEDGMENTS.} This work was supported by the State Fundamental Research Foundation of Ukraine (Grant No. F7/267-2001).
We used the LEDA database (http://leda.univ-lyon1.fr) and NED (http://nedwww.ipac.caltech.edu). 

{}
\end{document}